\documentclass{optica-article}
\journal{opticajournal} % for journals or Optica Open

\articletype{Research Article}

\usepackage{lineno}
\usepackage{graphicx}
\usepackage{subcaption}
\usepackage{amsmath}%,amssymb}
\usepackage{float}

\begin{document}

%\preprint{APS/123-QED}

\title{Quantum Sensing of Birefringence Beyond the Classical Limit with a Hyper-Squeezed SU(1,1) Interferometer}% Force line breaks with \\

\author{Samata Gokhale\authormark{1}, Netanel P. Yaish\authormark{1}, Michal Natan\authormark{1}, Saar Levin\authormark{1}, Yogesh Dandekar\authormark{2}, and Avi Pe'er\authormark{1,*}}

\address{\authormark{1}Department of Physics and BINA Center for Nanotechnology,\\ Bar-Ilan University, Ramat Gan 5290002, Israel}
\address{\authormark{2}Department of Physics, Bar-Ilan University, Ramat Gan 5290002, Israel}

%\email{avi.peer@biu.ac.il.com}

\email{\authormark{*}avi.peer@biu.ac.il}


\begin{abstract*} 
Quantum interferometric sensing plays a crucial role in a wide range of applications, including quantum metrology, quantum imaging, and quantum lithography, where minute phase shifts carry valuable physical information. The strength of quantum sensing lies in surpassing classical sensitivity limits, particularly through the use of quantum correlations and squeezing to suppress optical shot noise. Birefringence sensing is crucial for various applications, as it provides detailed information about the material's structure, stress, composition, and environmental conditions. We present an interferometric scheme for detecting unknown small birefringence beyond the shot-noise limit of sensitivity that leverages the hyper-squeezing (squeezed in both number-phase and polarization correlations) within a pair of polarized nonlinear SU(1,1) interferometers, coupled by the birefringence. Specifically, two pairs of crossed-polarization nonlinear media, both generate and measure two-mode quantum light that is squeezed and polarization-entangled. We present a complete theoretical analysis of the interferometer’s sensitivity to small birefringence under realistic conditions of gain and internal loss, illuminating the potential for enhancement of the sensitivity by 3-15dB in practical, real-world experiments (the exact achievable enhancement is governed solely by the loss).
\end{abstract*}

%\maketitle

\section{\label{sec:Intro}Introduction}

Birefringence is an optical property that reflects the anisotropy of materials, where the index of refraction depends on the polarization axis of the light (relative to the primary axes of the medium). Birefringence can either be intrinsic to the medium (due to the microscopic structure of the unit-cell of a crystal), or it can be induced by external factors, such as mechanical stress or temperature gradients in otherwise isotropic materials, such as glass or plastic, causing stress-induced birefringence or photoelasticity \cite{Howell_2012, Chandrasekhar:78, Zhu:03}. These small changes in the optical properties of a material are therefore a major target for optical sensing, since they reflect the mechanical/thermal stress of devices, and allow for predicting and preventing mechanical failures, like breaking, or damaging points \cite{Murphy-Erin, hsu_acoustical_1974}. Consequently, birefringence sensing is an important tool in mechanical \cite{Diaz:19} and civil \cite{Dong:09} engineering to evaluate structural integrity. Also, in material science \cite{Alvarez:11}, birefringence sensing is used to study stress distribution in glass, polymers, and composites, and in the semiconductor industry \cite{10.1063/1.2402906}, it allows to check residual stress in wafers and optical components.

Interferometric birefringence sensing \cite{Liu:05, Birnbaum:74, Cochran:92, amt-14-6213-2021} is a powerful method to detect such tiny changes of the index of refraction by measuring the induced phase shifts. For small birefringence (phase shift), traditional interferometers face the limit of classical interferometry set by the shot-noise \cite{ou2020quantum}. To overcome this, interferometric quantum sensing harnesses quantum properties of non-classical light, like quantum correlations and squeezing to enhance the sensitivity beyond the shot noise limit \cite{ou2020quantum, Lukens:16, Du:18, Anderson:17, Frascella:19, PhysRevLett.117.013001, PhysRevLett.119.223604}. Applying quantum interferometry to birefringence sensing will allow for extremely sensitive detection of stress-induced birefringence, beyond the classical limits.

For birefringence sensing beyond the shot noise limit, we suggest to modify the well-known quantum SU(1,1) interferometer configuration, composed of two phase-sensitive optical parametric amplifiers (OPAs) placed in series, with a phase shift between them. In SU(1,1) interferometers, interference results from the phase-sensitive response of the nonlinear OPAs: First, the SU(1,1) configuration automatically produces the necessary squeezing within the interferometer itself, and furthermore, an SU(1,1)-based sensor is very robust to detection inefficiency (as explained below), allowing to operate with simple, commercially available photo-detectors / cameras (as opposed to SU(2), where near-ideal detectors are necessary for sub-shot noise sensitivity).  

The phase sensitivity of an interferometer sensor is defined as the minimum detectable phase $\delta\varphi$, which is estimated according to : 
\begin{equation}\label{sensitivity}
    \delta\varphi^2 = \left(\frac{\Delta N}{\left|\frac{\partial \langle N \rangle}{\partial\varphi}|_{\varphi_0}\right|}\right)^2
\end{equation}
\cite{Xiaoying}, where the numerator $\Delta N = \sqrt{\langle N^2\rangle - \langle N \rangle^2}$ is the quantum fluctuations in the measured quantity (e.g., detected intensity(average number of photons)) and the denominator $\left|\frac{\partial \langle N \rangle}{\partial\varphi}|_{\varphi_0}\right|$ reflects the rate of change for the average intensity with respect to small phase shifts near $\varphi_0$ (the phase working point). Thus, the phase sensitivity can improve when the signal responds sharply to a phase change (large denominator) or when the measurement noise is small (small numerator). Using standard laser light, the phase sensitivity of the classical SU(2) interferometer is given by $\delta\varphi_{c}=\frac{1}{\sqrt{\langle N \rangle}}$ where \textit{N} is the total intensity detected during the measurement time. This is known as the classical or shot-noise limit (SNL). To improve an SU(2) interferometer beyond the SNL, one can inject properly prepared squeezed states into the unused input port, which leads to a reduction in the measurement noise below the SNL. However, to enjoy the resulting improvement in sensitivity, the photo-detectors must be near-ideal (quantum efficiency $\approx1$) and sufficiently low-noise to detect the reduced optical noise. Such detectors are technically demanding, or even unavailable in some ranges of the optical spectrum (like IR and UV). In contrast, the SU(1,1) interferometer does not reduce the noise (if at all, it enhances it), so the nominator remains large, but the slope of the response can be much larger, leading to an improved signal to noise ratio, and thereby sensitivity. Under ideal conditions, the phase sensitivity can approach the ultimate quantum limit $\varphi_H=\frac{1}{\langle N \rangle}$, called the Heisenberg limit ($H$), which provides a significant improvement for large intensity \cite{ou2020quantum,Lukens:16,Du:18,Anderson:17,Frascella:19,PhysRevLett.117.013001,PhysRevLett.119.223604}. 

Previously, the null interferometric method was used in \cite{Birnbaum:74} to measure stress-induced birefringence, where a birefringent sample is placed inside a Fabry-Perot interferometer. An electro-optic cell was used to cancel the sample birefringence by rotating its axis and tuning its voltage, which was then calibrated to yield the phase shift caused by the unknown birefringence. Another example used a Fizeau interferometer with a variable retarder and a non-polarizing beam splitter to measure stress-induced birefringence \cite{Cochran:92} with a known principal axis. 

\begin{figure}[htbp]
        \centering
       \includegraphics[width=1\linewidth]{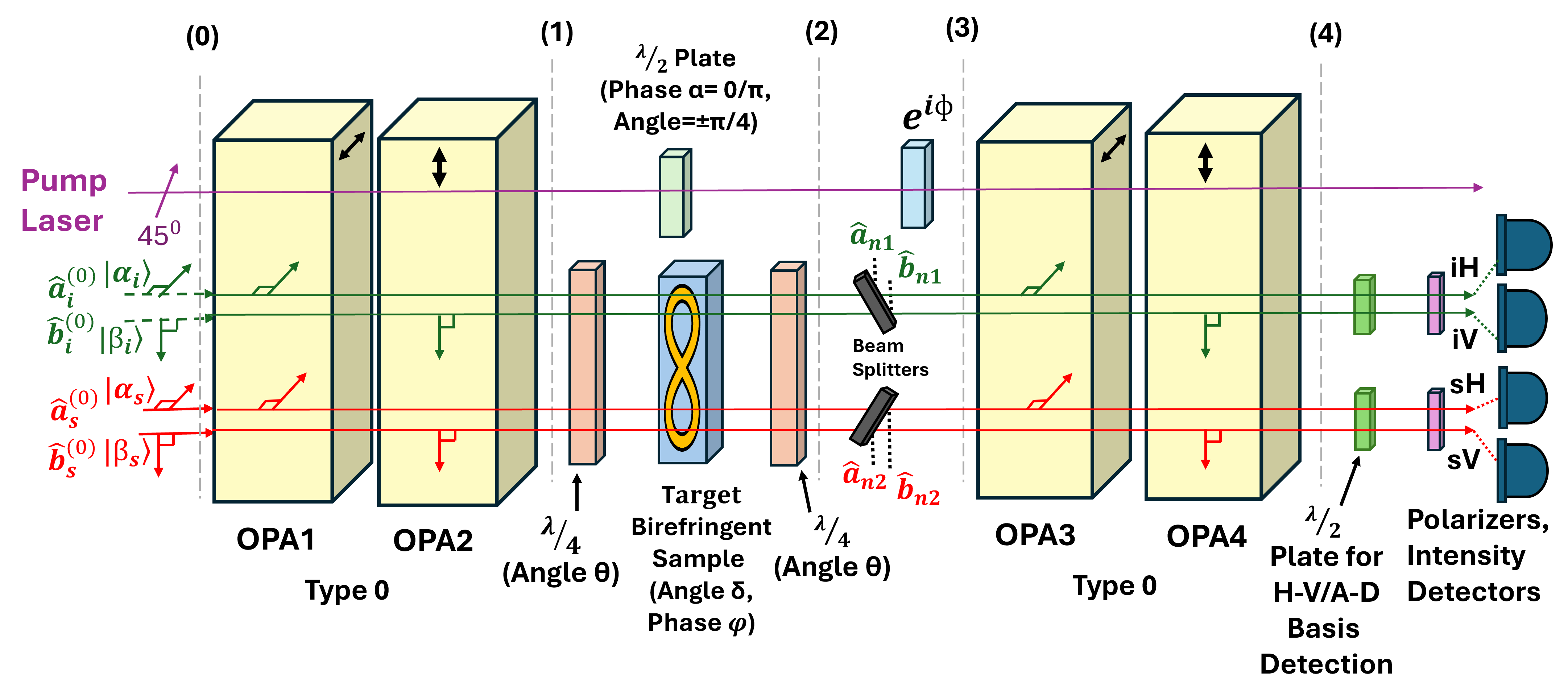}
        \caption{\textbf{Schematic layout of the proposed dual SU(1,1) interferometer for measuring birefringence.} Two pairs of OPAs are arranged in series (one for each polarization, $H$ and $V$) and the target birefringent sample in between. Quarter/half-wave plates are incorporated on the intermediate beam paths to manipulate the polarization of the pump (the detection basis) and the signal+idler (the polarization Bell state). The birefringent phase of the sample is $\varphi$, and its axis is at an angle $\delta$ with respect to the crystal axes of the OPAs. The vertical dashed lines (marked (0)...(4)) denote various intermediate planes in the configuration where the polarization state can be calculated. The figure shows the symmetric configuration, where the target birefringent sample is placed between two quarter-wave plates (at the same angle), but one can equally place the sample also before both wave-plates or after them. Our model can calculate sensitivity maps for all three cases (and all four Bell states.)}
       \label{Setup}
\end{figure}

We propose and analyze the configuration of figure \ref{Setup}, which consists of a dual SU(1,1) interferometer (one for each polarization horizontal ($H$) and vertical ($V$)) that are coupled with each other by an unknown birefringent plate (whose phase and principal axis are the sensing target). The dual SU(1,1) interferometer is composed of two pairs of polarization-selective optical parametric amplifiers (OPAs) that are arranged in series with the target birefringent sample in between. The interferometer input is stimulated by a strong classical coherent seed $|\alpha \rangle$ on the horizontal and seed $|\beta \rangle$ on vertical modes, and the output is detected directly by measuring the intensity (signal and/or idler) of each polarization independently. Each pair of OPAs has a crossed-axis configuration \cite{Paul_Kwiat}, such that each crystal in the primary pair squeezes the light in one polarization, whereas the crystals in the secondary pair measure these polarizations. When pumped at $45^o$, a pair of crossed-axis crystals generate the hyper-squeezed Bell state $| \Phi^+ \rangle = \frac{1}{\sqrt{2}}(|HH \rangle + |VV \rangle)$, which is both polarization-entangled and two-mode squeezed. By flipping the pump polarization from $45^o$ to $-45^o$ (with a $\lambda/2$ plate for the pump), we can also generate the $| \Phi^- \rangle = \frac{1}{\sqrt{2}}(|HH \rangle -|VV \rangle)$  Bell state. If we analyze the $| \Phi^- \rangle$ state in the diagonal - anti-diagonal basis ($A$-$D$) at $\pm45^o$ it transforms as $| \Phi^- \rangle\!=\!\frac{1}{\sqrt{2}}(|HH \rangle \!-\!|VV \rangle)\!=\! \frac{1}{\sqrt{2}}(|AD \rangle\!+\! | DA \rangle)\!=\!| \Psi^+ \rangle $, indicating that $| \Psi^+ \rangle$ can also be generated from the $| \Phi^- \rangle$ Bell state by simply rotating the polarization of the squeezed light by $45^o$ (with a $\lambda/2$ plate for the signal/idler) before the measurement OPAs (forming a time-reversed Hong-Ou-Mandel effect \cite{yaish2025generationdetectionhyperentangledbell, chen2007deterministic,chen2018polarization}).  This basis-dependent behavior reflects the polarization correlations between the modes that are encoded in the form of Bell states. Finally, one can also generate the singlet Bell state $| \Psi^- \rangle = \frac{1}{\sqrt{2}}(|HV \rangle - | VH \rangle)$ by flipping the phase of only one of the (signal/idler) intensity at only one polarization from the $| \Psi^+ \rangle$ state (with a frequency-selective wave-plate). All Bell states can be detected in H-V or A-D basis using a $\lambda/2$ plate placed after the second pair of OPAs. Also, this configuration operates in a fully collinear geometry, and relies on intensity measurements of the seeded, strongly correlated output fields in the high-power regime, rather than on post-selected two-photon coincidences. In our configuration, the second pair of OPAs act as “physical detectors” of the hyper-squeezed correlated polarizations in pairs either by amplifying or annihilating them, depending on the phase.

The quantum state of the light in the high-gain, seeded regime ($g=2$ in our simulation) is highly multi-photon and cannot be described as two-photon Bell-states of single photons. However, the presence of multi-photon pairs at such high gain does not imply the disappearance of polarization entanglement (or squeezing). Instead, the quantum correlations (squeezing) become distributed among the participating modes. The enhanced performance of birefringence sensing beyond the classical limit directly originates from these collective correlations and their nonlinear amplification within the SU(1,1) interferometer. 

In what follows, we describe the results of a complete theoretical analysis of this interferometric configuration that takes advantage of high-squeezing and polarization correlations to probe a small unknown birefringence beyond the classical shot noise limit (the full derivation of the model is provided in the Theoretical Model section). We calculate the interferometric sensitivity to small variations of the birefringent phase and explore its dependence on the various parameters of the interferometer. Specifically, we examine the dependence of the sensitivity at various working points of the interferometer in both initial phase of the target plate and the orientation angle of its primary axis. The resulting 2D maps also show regions of sub-shot-noise sensitivity at surprising, non-intuitive working points, which we discuss. We also explore the effect of both parametric gain and internal loss on the sensitivity, which are crucial in choosing the optimal conditions for sub-shot-noise sensing, showing that the improvement of the sensitivity with parametric gain is limited by the internal loss of the interferometer, as one would expect.

\section{\label{sec:Results}Results}
\begin{figure}[htbp]
        \centering
       \includegraphics[width=1\linewidth]{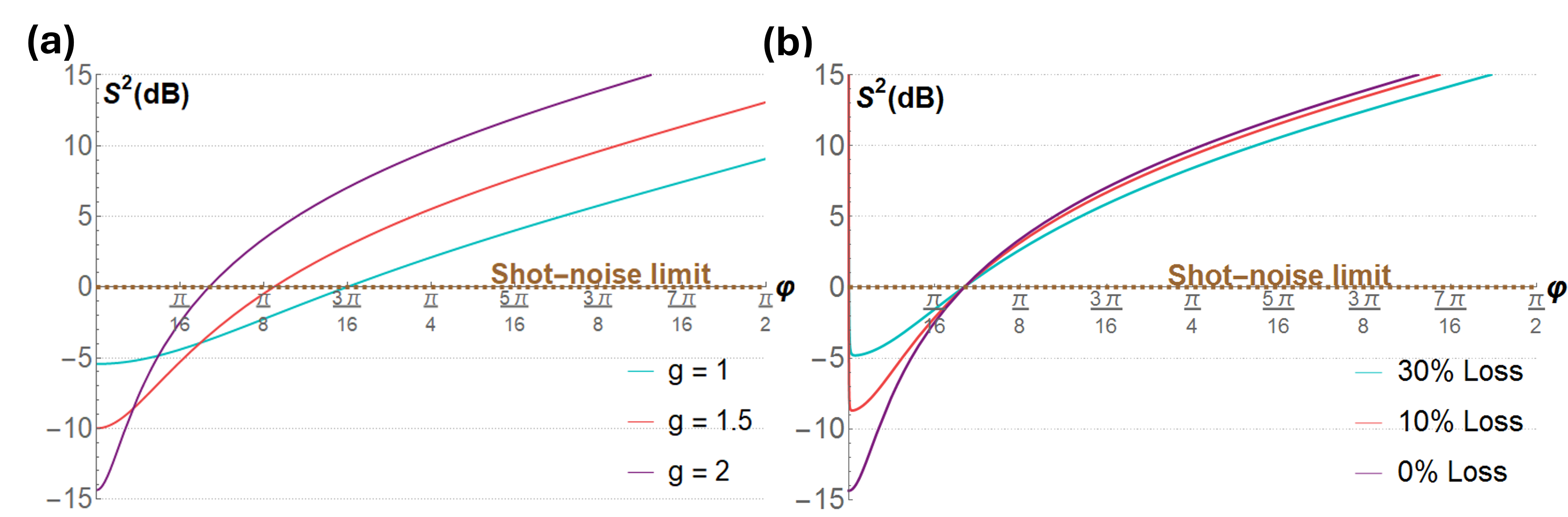}
        \caption{\textbf{Sensitivity vs parametric gain (a) and internal losses (b)}. Both graphs show the relative sensitivity squared $S^2=\left(\delta\varphi/\delta\varphi_c\right)^2$ as a function of the birefringent phase $\varphi$ for\textbf{ (a)} varying squeezing gain ($g=1,1.5,2$). When probing the sample at an angle $\delta = \pi/2$ with $ | \Phi^+ \rangle =\frac{1}{\sqrt{2}}(|HH \rangle+|VV\rangle)$ state and measuring the total number of idler horizontal ($iH$) intensity and seed at signal horizontal ($sH$), we observe enhanced sensitivity around $\varphi=0$. Sensitivity improvements of -5.44 dB for $g=1$, -9.98 dB for $g=1.5$, and -14.36 dB for $g=2$ are observed and \textbf{(b)} varying internal loss ($l=0\%, 10\%,30\%$).}
       \label{gain_losses}
\end{figure}
\begin{figure}[htbp]
        \centering
       \includegraphics[width=1\linewidth]{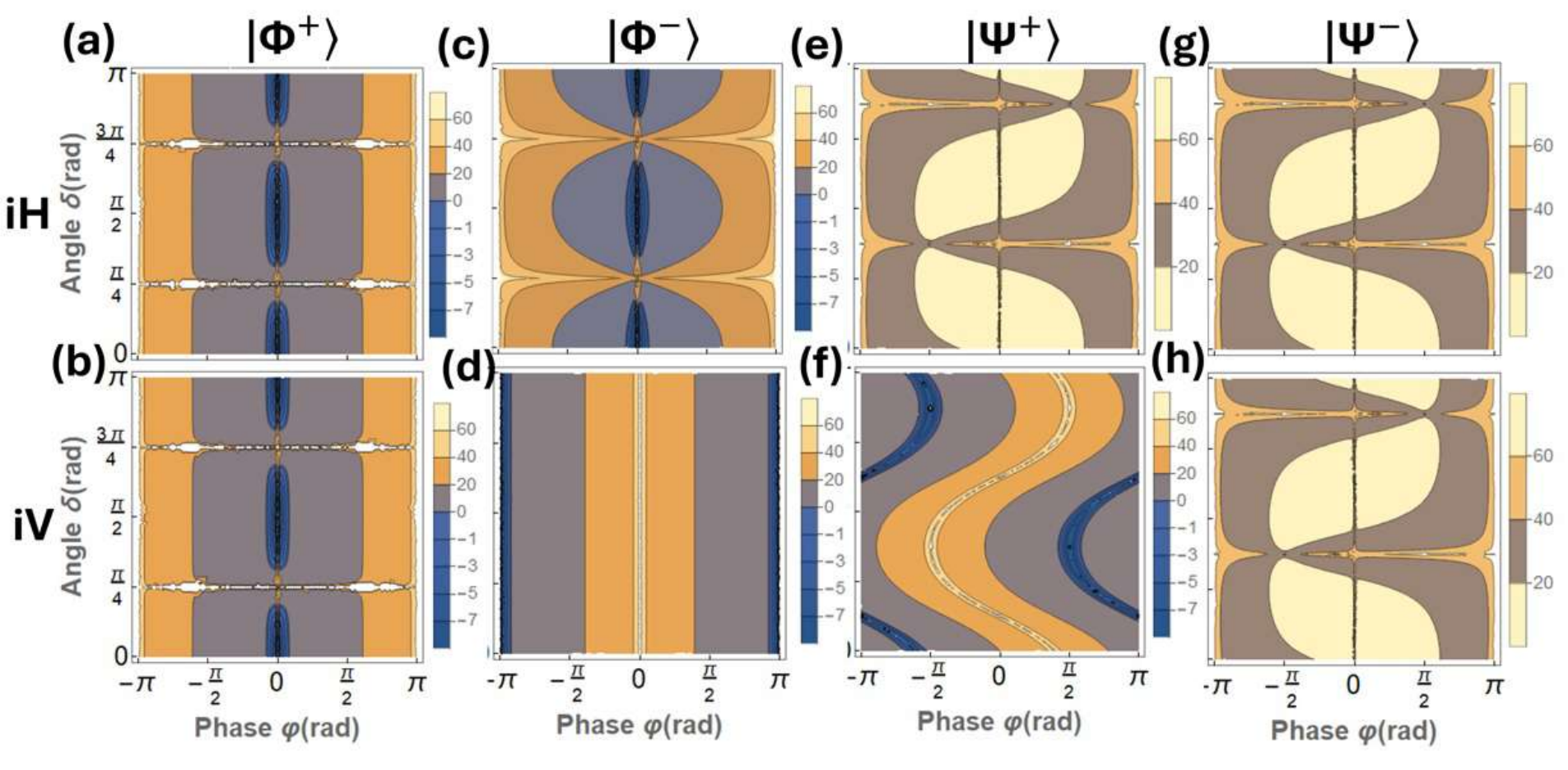}
        \caption{\textbf{Representative sensitivity maps for the different Bell states:} We present the relative sensitivity $S^2$ when the birefringent sample is placed symmetrically between the $\lambda/4$ plates, which are set in accordance with each one of the Bell states (according to table \ref{vis}) and for different seeding and detection polarizations measurement in $H$--$V$ basis: The columns represent the four Bell states and the rows represent seeding -- detection configurations, as follows - In the top (bottom) row we seed the signal at horizontal (vertical) polarization and detect the idler intensity at the same polarization. One can observe that for $ | \Phi^+ \rangle =\frac{1}{\sqrt{2}}(|HH  \rangle+| VV \rangle)$ \textbf{(a \& b)}, both polarizations show the same map (as expected from the symmetry of $| \Phi^+\rangle$) with quantum enhanced sensitivity along the $\varphi_0\!=\!0$ line, but excluding regions at $\delta=\pi/4, 3\pi/4$. For $ | \Phi^- \rangle =\frac{1}{\sqrt{2}}(|HH  \rangle-| VV \rangle)$ \textbf{(c \& d)}, the two polarizations show enhanced sensitivity along opposite lines; $\varphi_0=0$ for $H$ detection (excluding $\delta=\pi/4, 3\pi/4$), and $\varphi_0 = \pm\pi$ for $V$ detection (independent of the sample orientation $\delta$). For the $| \Psi^+ \rangle =\frac{1}{\sqrt{2}}(|HV \rangle+| VH \rangle) $ state \textbf{(e \& f)}, show no improvement in sensitivity for the $H$ detection, whereas $V$ detection shows a surprising "wiggly" striped pattern of improved sensitivity.For the singlet  $| \Psi^- \rangle =\frac{1}{\sqrt{2}}(|HV \rangle-| VH \rangle) $ state \textbf{(g \& h)}, shows the same behavior as \textbf{(e)} for $| \Psi^+ \rangle =\frac{1}{\sqrt{2}}(|HV\rangle+|VH \rangle) $ state. Internal losses of $10\%$ in intensity are assumed for all the maps.}
       \label{qwp_biref_qwp_plots}
\end{figure}

The calculated sensitivity to the birefringence phase depends on many parameters: the parametric gain $g$ of the OPAs, the losses within the interferometer, the measurement working point of the birefringence phase $\varphi$ and plate angle $\delta$ and the seed frequency (signal/idler) and polarization. We present the relative sensitivity in several cross sections: First, figure \ref{gain_losses} depicts the sensitivity for several values of the parametric gain and internal loss, showing that the gain dictates the maximum possible sensitivity enhancement, but the practically achieved enhancement is limited by the loss, as expected. In addition, figure \ref{qwp_biref_qwp_plots} presents 2D --maps of the relative sensitivity for parametric gain $g\!=\!2$ and $10\%$ loss (a representative value for a typical experimental configuration) as a function of the working point $(\varphi,\delta)$ for various configurations of seeded (measured) mode (in polarization) and for various locations for placing the plate, showing the different regions of sensitivity enhancement for various Bell states. For the sake of brevity, we show the maps for the symmetric configuration (sample between $\lambda/4$ plates) with a seed of horizontal (vertical) idler and measuring the horizontal (vertical) signal (see the caption for details). The maps for locations before (plane (1) in figure \ref{Setup}) and after (plane (2) in figure \ref{Setup}) the $\lambda/4$ plates, as well as seeding and detection in opposite polarization and in cross basis operation, are presented in the supplementary material. At $10\%$ losses, for parametric gain $g\!=\!2$, the 2D sensitivity maps show maximum improvement of birefringence sensitivity by 8.78 dB below the shot-noise limit (See Fig. \ref{gain_losses}(b)). 
Clearly, when the $| \Phi^+ \rangle =\frac{1}{\sqrt{2}} (\vert HH \rangle+| VV \rangle)$ state probes the sample, both polarizations exhibit the same 2D map of sensitivity to birefringence (Figure \ref{qwp_biref_qwp_plots} (a), (b)), as expected from an inherently symmetric state. However, for $| \Phi^- \rangle =\frac{1}{\sqrt{2}}( \vert HH \rangle-| VV \rangle)$ state, as $H$ and $V$ intensity are anti-correlated (out-of-phase), they exhibit out-of-phase behavior (Figure \ref{qwp_biref_qwp_plots} (c), (d)). Since $| \Psi^+ \rangle =\frac{1}{\sqrt{2}}(|HV\rangle+|VH \rangle) $ state is rotated through $\pi/8$ to $A$-$D$ basis but measured in $H$-$V$ basis, $H$ intensity do not show improvement in birefringence sensitivity, but $V$ intensity show improved birefringence sensitivity in "wiggly" striped pattern as the presence of the birefringent sample breaks the symmetry (Figure \ref{qwp_biref_qwp_plots} (e), (f)).

\section{\label{sec:level2}Theoretical Model}

For our configuration (Figure \ref{Setup}), we considered the quantum evolution of four modes, namely idler horizontal, idler vertical, signal horizontal and signal vertical (and two pump modes, taken as classical and undepleted). Through the process of parametric down conversion (PDC, either spontaneous or seeded), the OPA1(OPA2) with the horizontal (vertical) polarization axis generates only a horizontally (vertically) polarized signal and idler.

Hence, the horizontal (vertical) signal and idler field operators $a_{s,i}$($b_{i,s}$) after OPA1 and OPA2 can be written using the standard expression for OPA evolution, as  \\
\begin{equation} \label{OPA_eqaution}
\begin{split}
    \hat a^{(1)}_{i} &= (\hat a^{(0)}_{i} + \alpha_{i})\cosh g +  (\hat a^{(0)\dag}_{s}  + \alpha_{s})  \sinh g\\
    \hat a^{(1)}_{s} &= (\hat a^{(0)}_{s} +\alpha_{s})\cosh g + (\hat a^{(0)\dag}_{i} + \alpha_{i}) \sinh g\\
     \hat b^{(1)}_{i} &= (\hat b^{(0)}_{i} + \beta_{i})\cosh g + (\hat b^{(0)\dag}_{s} + \beta_{s}) \sinh g\\
    \hat b^{(1)}_{s} &= (\hat b^{(0)}_{s} + \beta_{s}) \cosh g + (\hat b^{(0)\dag}_{i}  + \beta_{i})\sinh g,
\end{split}
\end{equation}
where $g=\chi\alpha_p l$  is the parametric gain, assumed same for all OPAs ($\chi$ the nonlinear coefficient, $\alpha_p$ the pump amplitude, and $l$ the OPA length), $ a^{(0)}_{s,i}$ ($b^{(0)}_{s,i}$) are the input horizontal (vertical) field operators (acting on vacuum), a classical coherent seed $\alpha$ ($\beta$) on horizontal (vertical) modes. \footnote{We assume only single mode seeding here. Two mode seeding can be considered as well, but is not needed since it does not improve the sensitivity results.} The superscript of field operators denotes their location in the configuration, which is denoted out by dashed line planes in figure \ref{Setup}, the subscript denotes idler (i) or signal (s) intensity.

After the OPAs, these four modes then pass through 3 polarization elements - two quarter-wave plates ($\lambda/4$ plates) at an angle $\theta$ and the birefringent target with an angle $\delta$ and phase $\varphi$. We considered the sensitivity to a small birefringence when placing the target birefringent sample in one of three locations: before the first $\lambda/4$ plate (Plane 1) presented in supplemental material, between the two $\lambda/4$ plates presented in main text, and after the second $\lambda/4$ plate (Plane 2) presented in supplemental material. Since all three are simply birefringent windows (of different phases and angles), we use the Jones matrix of a general birefringent window to treat each one: 
\begin{equation}
    \begin{pmatrix}\label{theta}
\hat a^{(2)}_{i,s} \\ \hat b^{(2)}_{i,s}
\end{pmatrix}
=
\begin{pmatrix}
\cos \frac{\psi}{2}-i\cos2\gamma\sin \frac{\psi}{2}   & -i\sin2\gamma\sin \frac{\psi}{2} \\
 -i\sin 2\gamma \sin \frac{\psi}{2} &  \cos \frac{\psi}{2}+i\cos2\gamma\sin \frac{\psi}{2}
\end{pmatrix}
\begin{pmatrix}
\hat a^{(1)}_{i,s} \\ \hat b^{(1)}_{i,s}
\end{pmatrix}, 
\end{equation}
where $\gamma$ is the angle of the principal axis, and $\psi$ is the phase. For the $\lambda/4$ plates $\psi=\pi/2$ and the angle is $\gamma=\theta$ (the same for both $\lambda/4$ plates), and for the target birefringent sample, we assume a small unknown phase $\psi=\varphi$ and a tunable angle $\gamma=\delta$.  We can now propagate the field operators through all three elements by multiplying this Jones matrix three times with the proper angle and phase, and in the order of their setting. 

To account for losses within the interferometer, we introduce "fictitious" beam splitters (BS) with auxiliary vacuum inputs from their other ports \cite{michael2019squeezing, PhysRevLett.127.173603}, where the reflection  $r_{i,s}=\sqrt{1-{t}_{i,s}^2}$ represents the loss. 
 
To control the measurement basis at the second pair of OPAs according to the different Bell-states, we also manipulate at this stage the phases of the various polarization components of the signal/idler, as shown in figure \ref{Setup} between planes (2)--(3): $\phi$ is the overall SU(1,1) phase of between the pump and the idler and signal intensity of both polarizations. To consider measurement at the singlet Bell-state $| \Psi^- \rangle$, a sign-flip ($\beta=\pi$) is introduced to only one polarization component of only the signal intensity, as discussed before. 

The transformation of the field operators due to the losses and those various phases is:
\begin{equation}
\begin{split}
\begin{pmatrix}\label{beta}
\hat a^{(3)}_{i} \\ \hat b^{(3)}_{i}
\end{pmatrix}
&=
\begin{pmatrix}
e^{i\phi}\left[t_i\hat a^{(2)}_{i} +  r_i\hat a_{n1}\right]\\ e^{i\phi} \left[t_i\hat b^{(2)}_{i} +r_i\hat b_{n1}\right]
\end{pmatrix}\\
    \begin{pmatrix}
\hat a^{(3)}_{s} \\ \hat b^{(3)}_{s}
\end{pmatrix}
&=
\begin{pmatrix}
t_s\hat a^{(2)}_{s} + r_s\hat a_{n2}\\e^{-i\beta}\left[t_s\hat b^{(2)}_{s} +r_s\hat b_{n2}\right]
\end{pmatrix},
\end{split}
\end{equation}
where $t_{i,s} (r_{i,s})$ is the amplitude transmission (reflection) of the idler, signal modes of the fictitious beam splitters, and $\hat a_{n1,2}$, $\hat b_{n1,2}$ are the auxiliary vacuum modes (losses) that couple with propagating quantum field operators. Furthermore, to generate the $|\Phi^-\rangle$ Bell state, the pump polarization is flipped from angle $\alpha=+45^0$ to $\alpha=-45^0$ using a $\lambda/2$ plate on pump path. This is equivalent to the evolution of parametric gains (${g1, g2}$) to (${h1, h2}$) because of the pump $\lambda/2$ plate as expressed below:
\begin{equation}
     \begin{pmatrix}\label{alpha}
h1 \\h2
\end{pmatrix}
=
\begin{pmatrix}
\cos 2\alpha   & \sin 2\alpha \\
 \sin 2\alpha &  -\cos 2\alpha
\end{pmatrix}
\begin{pmatrix}
g1 \\ g2
\end{pmatrix}, 
\end{equation}
The values of $\alpha, \beta, \theta$ to generate the four Bell states are tabulated in the Table \ref{vis}.
\begin{table}[h!]
\centering
\begin{tabular}{|c c c c|} 
 \hline
Bell State & $\alpha$ & $\beta$ & $\theta$  \\ [0.5ex] 
 \hline
 $\frac{1}{\sqrt{2}} (\vert HH \rangle+| VV \rangle)$ & $\pi/4$ & 0 & 0 \\ 
 $\frac{1}{\sqrt{2}} (\vert HH \rangle-| VV \rangle)$ & 0& 0 & 0 \\ 
 $\frac{1}{\sqrt{2}} (\vert HV \rangle+| VH \rangle)$ &  0 & 0 & $\pi/8$ \\ 
 $\frac{1}{\sqrt{2}} (\vert HV \rangle-| VH \rangle)$ &  0 & $\pi$ & $\pi/8$ \\  [1ex] 
 \hline
\end{tabular}
\caption{Values  of $\theta$ in equation \eqref{theta}, values of $\beta$  in equation \eqref{beta} and values of $\alpha$ in equations \eqref{alpha} to generate four Bell states with the setup of figure \ref{Setup}.}
\label{vis}
\end{table}
Finally, we reach the measurement stage, where these field operators pass through the OPA3 and OPA4 (according to Equation \eqref{OPA_eqaution} once more), forming the output operators {$ \hat a^{(4)}_{i,s}, \hat b^{(4)}_{i,s} $} of the dual SU(1,1) interferometer. At detection, only idler horizontal (vertical) or total idler + signal horizontal (vertical) intensity can be detected. The total $H$ and $V$ intensity detection according to: 
\begin{equation}
\begin{split}
   \langle N_{H} \rangle &= a^{(4)\dag}_{i} a^{(4)}_{i} + a^{(4)\dag}_{s} a^{(4)}_{s} \\
    \langle N_{V} \rangle & = b^{(4)\dag}_{i} b^{(4)}_{i} + b^{(4)\dag}_{s} b^{(4)}_{s}
\end{split},\label{photonnumber}
\end{equation}
and their quantum fluctuations $\left ( \Delta N_{H,V} = \sqrt{\langle N_{H,V}^2\rangle - \langle N_{H,V} \rangle^2} \right)$ can be calculated, dependent on all the parameters above of the interferometer ($\alpha, \beta, \theta, \phi,\varphi$ and $ \delta$).

We can now turn to the sensitivity of measuring birefringence and calculate it (using equation \eqref{sensitivity} and the measured $N_{H}$ and $N_{V}$) in terms of the working point of the interferometer (birefringent phase $\delta$ and angle $\gamma$ of the target sample). To evaluate the improvement in quantum sensitivity of SU(1,1) over the classical shot-noise limit, we normalized the squared quantum sensitivity of SU(1,1) by the squared shot-noise limit and then expressed the resulting expressions separately for $H$ and $V$-- polarized light in decibels (dB); the resulting normalized sensitivity is denoted as $S^2$. We calculated the shot-noise limit from $1/\sqrt{N}$ where $N$ the total intensity from all four modes) traversed till the plane (3) in figure \ref{Setup}. 

\section{Conclusions}
A seeded SU(1,1) interferometer is an "ideal sensor" of phase below the classical shot noise limit because it alleviates the need for low noise, highly efficient detectors, and since it automatically generates quantum squeezing within the interferometer with an inherently stable phase. We performed a complete theoretical analysis of birefringence sensing using two polarization-sensitive SU(1,1) interferometers that are coupled by a birefringent sample. Our interferometer both generated and measured two-mode signal -- idler fields that are hyper-squeezed (correlated in polarization and two-mode squeezed in amplitude-phase). This scheme leverages the advantages of SU(1,1) interference in both two-mode squeezing and polarization correlations to obtain phase-sensitivity below the shot-noise limit. Our results show a clear quantum enhancement in birefringence sensitivity for all the Bell states, with a rich structure in terms of the configuration parameters: seeding mode and polarization, location and orientation of the birefringent sample, phase working point, and measurement choice (mode and polarization basis). This analysis therefore, expands the utility of SU(1,1) sensors to the regime of hyper-squeezing (squeezed in both number-phase and polarization correlations, in our case).  

\begin{backmatter}
\bmsection{Disclosures}
The authors declare no conflicts of interest.

\bmsection{Data availability} No data were generated or analyzed in the presented research
\end{backmatter}

\bibliography{biref_sensing}% Produces the bibliography via BibTeX.

\end{document}